\documentclass[aps,pre,showpacs,amsmath,amssymb,reprint]{revtex4-1}

\usepackage{bm}
\usepackage{graphicx}
\usepackage{color}

\begin{document}

\title{Identification of Separation Wavenumber between Weak and Strong Turbulence Spectra
for Vibrating Plate}

\author{Naoto Yokoyama}
\email{yokoyama@kuaero.kyoto-u.ac.jp}
\affiliation{Department of Aeronautics and Astronautics, Kyoto University, Kyoto 615-8540, Japan}

\author{Masanori Takaoka}
\email{mtakaoka@mail.doshisha.ac.jp}
\affiliation{Department of Mechanical Engineering, Doshisha University, Kyotanabe 610-0394, Japan}

\date{\today}

\begin{abstract}
A weakly nonlinear spectrum and a strongly nonlinear spectrum
coexist in a statistically steady state of elastic wave turbulence.
The analytical representation of the nonlinear frequency is obtained
by evaluating the extended self-nonlinear interactions.
The {\em critical\/} wavenumbers
at which the nonlinear frequencies are comparable with the linear frequencies
agree with the {\em separation\/} wavenumbers between the weak and strong turbulence spectra.
We also confirm the validity of our analytical representation of the separation wavenumbers
through comparison with the results of direct numerical simulations
by changing the material parameters of a vibrating plate.
\end{abstract}

\pacs{62.30.+d, 05.45.-a, 46.40.-f}

\maketitle

\section{Introduction}
The coexistence of weakly nonlinear and strongly nonlinear spectra,
which is called critical balance,
has been predicted
in several {\em anisotropic\/} wave turbulence systems~\cite{goldreich1995toward,*FLM:409533,*nazarenko3488critical}.
The characteristic time of the linear regime is the wave period:
e.g.,
the Alfv\'en wave period in magnetohydrodynamics,
axial component of the rotation frequency in rotating turbulence,
and horizontal component of the buoyancy frequency in stratified turbulence.
The nonlinear time scale
is the horizontal eddy turnover time in these three types of turbulence.
The critical balance is a plausible conjecture
that illustrates the energy flux in an anisotropic wave turbulence,
where the nonlinear time is comparable 
with the linear dispersion time across a wide range of wavenumbers.
Thus,
it is of great importance to understand the physics
in the wavenumbers at which the linear and nonlinear time scales are comparable.
Even if a system is {\em isotropic\/}, 
the linear time scale may be comparable with the nonlinear time scale.
In this paper,
the clarification of the coexistence in an isotropic wave turbulence system
is addressed from the viewpoint of the time scales.

Various energy spectra have been observed for the wave turbulence
in a thin elastic plate.
The possibility of the Kolmogorov--Zakharov spectrum, 
which is predicted by the weak turbulence theory~\cite{zak_book},
provides the motivation for these studies.
In a numerical study~\cite{during2006weak},
the weak turbulence spectrum $\mathcal{E}(k) \propto k [\log(k_{\ast}/k) ]^{1/3}$ is obtained as a statistically steady spectrum,
where $k$ is the magnitude of the two-dimensional wavenumber vector $\bm{k}$, i.e., $k=|\bm{k}|$.
Contrarily,
energy spectra such as $\mathcal{E}(k) \propto k^{-0.2}$ and $k^{0}$
are experimentally observed~\cite{boudaoud2008observation,mordant2008there,PhysRevLett.107.034501}.
Moreover,
the existence of the spectra with the energy cascade $\mathcal{E}(k) \propto k^{-1}$ and with the wave action cascade $\mathcal{E}(k) \propto k^{-1/3}$
is predicted by the dimensional analysis~\cite{nazarenkobook}.
The energy spectrum obtained by the numerical simulation is the steady solution of the kinetic equation of the weak turbulence theory,
while the energy spectra obtained in the experiments and by the dimensional analysis
are outside the scope of the weak turbulence theory.
In our earlier study~\cite{PhysRevLett.110.105501},
we provided
a unified perspective on the variability of the spectra
by performing direct numerical simulations using the F\"{o}ppl--von K\'{a}rm\'{a}n equation and found that
the strength of the nonlinearity causes the variability.

Recently,
it was shown that the experimentally observed spectrum
may be explained by the strong dissipation in the small frequencies~\cite{0295-5075-102-3-30002}.
The boundary conditions and homogeneity of the plate 
may also affect the energy spectra, as pointed out in Ref.~\cite{PhysRevE.84.066607}.
We are interested in the universal dynamics in the inertial subrange,
in which the external force and dissipation do not affect the dynamics.
The subject concerned in this manuscript is
an extension of the research~\cite{during2006weak},
whose attractive subtitle is ``Can one hear a Kolmogorov spectrum?''.

\begin{figure}[]
\includegraphics[scale=.9]{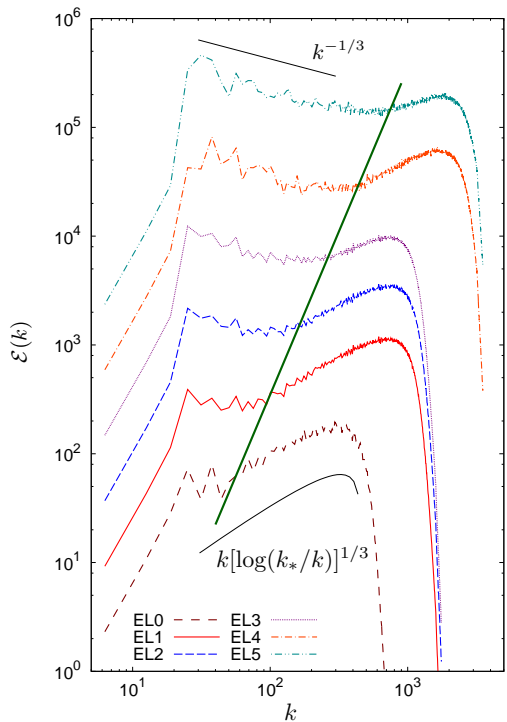}
 \caption{(Color online)
 Energy spectra at various energy levels, EL0 to EL5, in statistically steady states.
 The solid line indicates the separation between the weak and strong turbulence spectra.
 The weak turbulence spectrum $k [\log(k_{\ast}/k)]^{1/3}$,
 where $k_{\ast} = 144\pi$,
 and a power-law spectrum $k^{-1/3}$
 are shown for reference.
 \label{fig:motivation}
 }
\end{figure}

In Ref.~\cite{PhysRevLett.110.105501},
the weak turbulence spectrum was obtained at low energy levels, 
while the strong turbulence spectrum,
which is similar to the spectrum of the wave action cascade,
was obtained at high energy levels.
Moreover, the coexistence of the spectra,
in which the weak and strong turbulence spectra, respectively, appear at small and large wavenumbers,
was obtained at intermediate energy levels.
The numerical simulations with larger mode numbers (Fig.~\ref{fig:motivation}) show that
if the inertial subrange is large enough,
the coexistence state is generally achieved.
This generality was not found in Fig.~2 of Ref.~\cite{PhysRevLett.110.105501}.
In Fig.~\ref{fig:motivation},
the {\em separation\/} wavenumbers between the two types of the spectra lie on a straight line,
which suggests that there exists a simple law.

A similar bending structure of the energy spectra
is numerically observed using another type of external force and dissipation~\cite{PhysRevLett.111.054302}.
This fact suggests that the bending structure is universal.
The weak turbulence spectra at large wavenumbers
are commonly observed in the numerical simulations~\cite{during2006weak,PhysRevLett.110.105501,PhysRevLett.111.054302}.
It should be possible to estimate the separation wavenumbers by
balancing the linear and nonlinear time scales
owing to the following reason:
the weak turbulence theory is not applicable in the wavenumber range
in which the nonlinear time scales of the energy transfer are comparable with or shorter than the linear time scales~\cite{newell01,yokoyama2011wave}.
Note that
the linear time scale is comparable with the nonlinear time scale at a certain scale
in the statistically isotropic system of the elastic waves;
this is distinct from the comparability of the time scales over a wide range of wavenumbers in the systems reported as the critical balance.

In the experiment of a thin elastic plate~\cite{mordant2010fourier},
the nonlinear dispersion relation obtained from the space-time Fourier spectra
indicates a scaling law that is a function of the energy flux.
However, the coexistence of the weak and strong turbulence spectra
as well as the balance between the linear and nonlinear time scales
was not investigated.
The nonlinear time scales are usually estimated
by the nonlinear-interaction times
according to the weak turbulence theory in Refs.~\cite{newell01,Biven200128,Biven200398}.
However, because the unsteadiness of a system is required for
the finiteness of the nonlinear-interaction times, 
the nonlinear-interaction times cannot be used 
in statistically steady states.
Therefore, in the present analysis,
the nonlinear time scales are evaluated using the nonlinear frequency shifts,
which are the difference between the linear and nonlinear frequencies,
instead of the nonlinear-interaction times.
The nonlinear frequencies are derived 
by considering self-nonlinear interactions
in Refs.~\cite{newell01,Biven200128,Biven200398}.
The present system, however, requires an extension of this derivation
since it has the $1 \leftrightarrow 3$ and $3 \leftrightarrow 1$ interactions 
as well as the $2 \leftrightarrow 2$ interactions.
The {\em companion\/} elementary wave,
which has the wavenumber with the opposite sign,
should be taken into account
for the weakly nonlinear expansion to be consistent in our derivation. 
The interactions are referred to as ``extended self-nonlinear interactions'' in this paper.

In this paper,
we propose the analytical representation to evaluate the separation wavenumbers
from the viewpoint of the time scales.
First,
we show that
the nonlinear frequencies estimated by the extended self-nonlinear interactions
agree well with the representative nonlinear frequencies obtained from 
the time series in our simulations.
Then,
we show that 
the {\em critical\/} wavenumbers
at which the frequency shift in the nonlinear frequency
is comparable with the linear frequency
agree well with the separation wavenumbers.
The validity of our analytical representation is confirmed
by other series of numerical simulations 
in which the material parameters of elastic plates are changed.

\section{Formulation of Elastic Waves}
\subsection{Fourier representation}

The dynamics of the elastic waves propagating in a thin plate
is described by the F\"{o}ppl--von K\'{a}rm\'{a}n (FvK) equation~\cite{llelasticity}.
Under the periodic boundary condition,
the FvK equation is written
using the Fourier coefficient of the displacement $\zeta_{\bm{k}}(t)$
and that of the momentum $p_{\bm{k}}(t)$
as follows:
\begin{subequations}
\begin{align}
 \frac{d}{d t} p_{\bm{k}}
 &=  -\frac{Eh^2}{12 (1 -\sigma^2)} k^4 \zeta_{\bm{k}}
 \nonumber\\
 & \quad
- \frac{E}{2} 
\!\!
 \sum_{\bm{k}=\bm{k}_1+\bm{k}_2+\bm{k}_3}
\!\!\!\!\!\!\!\!\!\!\!
 \frac{|\bm{k} \times \bm{k}_1|^2|\bm{k}_2 \times \bm{k}_3|^2}{|\bm{k}_2+\bm{k}_3|^4}
 \zeta_{\bm{k}_1}\zeta_{\bm{k}_2}\zeta_{\bm{k}_3}
 ,
 \\
 \frac{d}{dt} \zeta_{\bm{k}} &= \frac{p_{\bm{k}}}{\rho}
,
\label{eq:fvkzetapb}%
\end{align}%
\label{eq:fvkzetap}%
\end{subequations}%
where $E$, $\sigma$, $\rho$, and $h$ are
the Young's modulus, Poisson ratio, density, and thickness of an elastic plate,
respectively.

The FvK equation~(\ref{eq:fvkzetap}) can be rewritten for the complex amplitude
\begin{align}
 a_{\bm{k}}(t) = 
\frac{\rho \omega_{\bm{k}} \zeta_{\bm{k}}(t) + i p_{\bm{k}}(t)}{\sqrt{2 \rho \omega_{\bm{k}}}}
,
\label{eq:defa}
\end{align}
as
\begin{align}
 \frac{da_{\bm{k}}}{dt} 
=& - i \omega_{\bm{k}} a_{\bm{k}}
\nonumber\\
&
- \frac{iE}{8\rho^2}
\!\!
 \sum_{\bm{k}=\bm{k}_1+\bm{k}_2+\bm{k}_3}
\!\!\!\!\!\!\!\!
\frac{|\bm{k} \times \bm{k}_1|^2|\bm{k}_2 \times \bm{k}_3|^2}{|\bm{k}_2+\bm{k}_3|^4 }
\nonumber\\
& \quad \times
\frac{
(a_{\bm{k}_1} + a_{-\bm{k}_1}^{\ast})
(a_{\bm{k}_2} + a_{-\bm{k}_2}^{\ast})
(a_{\bm{k}_3} + a_{-\bm{k}_3}^{\ast})
}{\sqrt{\omega_{\bm{k}}\omega_{\bm{k}_1}\omega_{\bm{k}_2}\omega_{\bm{k}_3}}}
.
\label{eq:fvka}
\end{align}
If the nonlinear terms were absent,
$a_{\bm{k}}$ would rotate clockwise in the phase space with the linear frequency
\begin{align}
 \omega_{\bm{k}} = \sqrt{\frac{Eh^2}{12(1-\sigma^2)\rho}} \, k^2
.
\label{eq:lineardispersion}
\end{align}
Thus,
the complex amplitude $a_{\bm{k}}$ represents
the elementary wave with the wavevector $\bm{k}$
and plays a central role in the weak turbulence theory.
The azimuthally integrated energy spectrum is defined as
$\mathcal{E}(k) = (\Delta k)^{-1} \sum_{k-\Delta k/2 \leq |\bm{k}| < k+\Delta k/2}
 \omega_{\bm{k}} |a_{\bm{k}}|^2$,
where $\Delta k$ is the bin width used to obtain the spectrum.

\subsection{Formulation of simulation}

Direct numerical simulations (DNS) were performed using the Fourier-spectral method
according to Eq.~(\ref{eq:fvka}).
To obtain statistically steady states,
we added the external force $\mathcal{F}_{\bm{k}}$,
which is effective at small wavenumbers,
and dissipation $\mathcal{D}_{\bm{k}}$, which is effective at large wavenumbers,
and the governing equation can be rewritten as
\begin{align}
\frac{da_{\bm{k}}}{dt} = -i\omega_{\bm{k}} a_{\bm{k}} + \mathcal{N}_{\bm{k}}
+\mathcal{F}_{\bm{k}}+\mathcal{D}_{\bm{k}}
.
\label{eq:fvkeq-a}
\end{align}
Here,
$\mathcal{N}_{\bm{k}}$ symbolically expresses the four-wave nonlinear interactions in Eq.~(\ref{eq:fvka}).

The external force $\mathcal{F}_{\bm{k}}$ is provided
so that
the absolute values of the complex amplitudes in the forced wavenumbers $|\bm{k}| \leq 8\pi$ are kept constant.
The dissipation $\mathcal{D}_{\bm{k}}$ is provided 
as the eighth-order hyperviscosity to achieve larger inertial subranges.
The pseudo-spectral method with the aliasing removal by the $4/2$ law
is used to obtain $\mathcal{N}_{\bm{k}}$
under the periodic boundary condition.
The numbers of the Fourier modes are
$N_{\mathrm{mode}}^2=512^2$ ($256^2$ alias-free) for EL0,
$N_{\mathrm{mode}}^2=1024^2$ ($512^2$ alias-free) for EL1--EL3, and
$N_{\mathrm{mode}}^2=2048^2$ ($1024^2$ alias-free) for EL4 and EL5.
The fourth-order Runge-Kutta method with a time step of $10^{-7}$s is used for the time integration. 
The time step should be small enough to resolve the fast dynamics of the large wavenumbers,
which is governed by the linear dispersion relation~(\ref{eq:lineardispersion}),
to reproduce the energy transfer due to the resonant interactions.
The parameters of the steel plate
which are shown in Ref.~\cite{boudaoud2008observation} were adopted
as the standard values.
The details of the numerical method are explained in
Ref.~\cite{PhysRevLett.110.105501}.

By changing the magnitude of the external force $\mathcal{F}_{\bm{k}}$,
the total energy in the system can be controlled.
The energy spectra are shown in Fig.~\ref{fig:motivation}.
Three energy levels, EL1, EL2, and EL3, were selected as the representatives
of the simulations to support our analytical results
because of the limited amount of computational time.
The ratio of the largest time scale to the smallest one
is of the order $N_{\mathrm{mode}}^2 \sim 10^6$
because of the linear frequency~(\ref{eq:lineardispersion}).

\section{Strategy for identifying separation wavenumber}

Let us find an analytical representation 
to identify the separation wavenumber
between the weak and strong turbulence spectra.
Our analysis begins with the fact that
the spectra at large wavenumbers represent the
weak turbulence spectra which are consistent with the weak turbulence theory.
For the weak turbulence theory to be applicable,
the linear time scale should be much smaller than the nonlinear time scale.
Therefore,
the separation wavenumber
is expected to be the critical wavenumber
at which the nonlinear time scale is comparable with the linear time scale.

In the studies of unsteady wave turbulence,
the nonlinear time scale is often estimated
using the nonlinear-interaction time $n_{\bm{k}}/(dn_{\bm{k}}/dt)$,
where $n_{\bm{k}}$ is the wave action
defined as
$n_{\bm{k}} \delta_{\bm{k},\bm{k}^{\prime}} = \langle a_{\bm{k}} a_{\bm{k}^{\prime}}^{\ast} \rangle$.
(See for example Refs.~\cite{newell01,Biven200128}.)
It should be noted that 
$n_{\bm{k}}/(dn_{\bm{k}}/dt)$ has a finite value only in a statistically unsteady state,
although it describes the time scale of the net energy transfer.
One might think that 
the self-similarity and the dimensional analysis
can be used for the applicability-limit scale for the weak turbulence theory
as is conventionally done~\cite[e.g.,][]{newell01,Biven200128,Connaughton200386}.
This idea is based on the fact that 
$n_{\bm{k}}/(dn_{\bm{k}}/dt)$, whose power-law dependence on wavenumbers
is different from that of the linear dispersion,
must cross over the linear time scale.
However, 
the crossover scale depends not only on the power-law exponents of the time scales with respect to the wavenumbers
but also on their coefficients.
Although the power-law dependence of $n_{\bm{k}}/(dn_{\bm{k}}/dt)$
on the wavenumbers can be obtained
by using the self-similarity and/or the dimensional analysis,
the coefficient cannot be found within this framework.
Furthermore, it cannot remain finite in the statistically steady states.

The energy flux $P$ is generally used by employing the relation
$d \mathcal{E}(k)/dt = - dP/dk$
for steady states of single turbulence~\cite{Connaughton200386}.
However, it is not appropriate to use $P$
in our present purpose to estimate the separation scale
between the two types of turbulence,
although $P$ is used in Ref.~\cite{mordant2008there}.
It is difficult to expect that a single value of $P$ can represent
the energy flux over the entire range of wavenumbers in the present system
since the turbulence shows different characteristics
on both sides of the separation wavenumber.
Furthermore,
since the Kolmogorov--Zakharov spectrum has the same self-similarity
as the Rayleigh--Jeans equilibrium spectrum,
which corresponds to the energy equipartition,
the energy flux is zero at the first order analysis of the weak turbulence theory~\cite{during2006weak}.

Our strategy adopted here is that
the nonlinear time scales are evaluated
not by the energy transfer 
but instead by the frequency shift due to the extended self-nonlinear interactions.
This implies that it is independent of the weak turbulence theory,
which is a statistical theory;
however, it depends on the weakly nonlinear expansion.
In this sense,
our estimation for the nonlinear time scale
is different from that in Refs.~\cite{newell01,Biven200398,Connaughton200386},
which is estimated by the resonant interactions.
Then,
the frequency shift
is balanced to the linear frequency
to estimate the separation wavenumber.
In fact,
this balance corresponds to equating the magnitudes 
of the linear and nonlinear terms 
on the right-hand side of Eq.~(\ref{eq:fvka}).

\section{Nonlinear Frequency}

\subsection{Estimation by extended self-nonlinear interactions}

Let us estimate the nonlinear frequencies to evaluate the nonlinear time scales.
Among the nonlinear interactions,
the combination of the wavenumbers, 
including the self-nonlinear interactions
in which one of $\bm{k}_1$, $\bm{k}_2$, and $\bm{k}_3$ 
appearing in Eq.~(\ref{eq:fvka}) is $\bm{k}$,
is confined to
$
(\bm{k}, \bm{k}_1, \bm{k}_2, \bm{k}_3) =
(\bm{k}, \bm{k}^{\prime}, \bm{k}, -\bm{k}^{\prime})
$
or
$
(\bm{k}, \bm{k}^{\prime}, -\bm{k}^{\prime}, \bm{k})
.
$
These self-nonlinear interactions are dominant if the nonlinearity is weak.
Then, the FvK equation~(\ref{eq:fvka}) can be rewritten as
\begin{align}
\frac{d}{dt} a_{\bm{k}}
 &= - i \omega_{\bm{k}} a_{\bm{k}}
-i \omega_{\bm{k}}^{\mathrm{s}}
(a_{\bm{k}} + a_{-\bm{k}}^{\ast}) 
 + \mathcal{N}_{\bm{k}}^{\mathrm{nn}}
,
\label{eq:FvKself}
\end{align}
where 
\begin{align}
 \omega_{\bm{k}}^{\mathrm{s}}
 &= 
 \frac{E}{4\rho^2}
 \sum_{\bm{k}^{\prime}}
\frac{|\bm{k} \times \bm{k}^{\prime}|^4}{|\bm{k} - \bm{k}^{\prime}|^4}
\nonumber\\
& \qquad \times
\frac{
|a_{\bm{k}^{\prime}}|^2 + |a_{-\bm{k}^{\prime}}|^2
 + a_{\bm{k}^{\prime}} a_{-\bm{k}^{\prime}} + a_{\bm{k}^{\prime}}^{\ast} a_{-\bm{k}^{\prime}}^{\ast}}%
{\omega_{\bm{k}}\omega_{\bm{k}^{\prime}}}
\nonumber\\
&
=
 \frac{E}{2\rho\omega_{\bm{k}}}
 \sum_{\bm{k}^{\prime}}
\frac{|\bm{k} \times \bm{k}^{\prime}|^4}{|\bm{k} - \bm{k}^{\prime}|^4}
|\zeta_{\bm{k}^{\prime}}|^2
.
\label{eq:freqshift}
\end{align}
Equation~(\ref{eq:freqshift}) represents the frequency increment from $\omega_{\bm{k}}$ 
due to the self-nonlinear interactions.
Note that $\omega_{\bm{k}}^{\mathrm{s}}$ is non-negative by definition.
The nonself-nonlinear interactive term $\mathcal{N}_{\bm{k}}^{\mathrm{nn}}$
is defined as
\begin{align*}
\mathcal{N}_{\bm{k}}^{\mathrm{nn}}
=&
- \frac{iE}{8\rho^2}
\!\!
 \sum_{\bm{k}=\bm{k}_1+\bm{k}_2+\bm{k}_3,
\bm{k}_1, \bm{k}_2, \bm{k}_3 \neq \bm{k}
}
\!\!\!\!\!\!\!\!
\frac{|\bm{k} \times \bm{k}_1|^2|\bm{k}_2 \times \bm{k}_3|^2}{|\bm{k}_2+\bm{k}_3|^4 }
\nonumber\\
& \quad \times
\frac{
(a_{\bm{k}_1} + a_{-\bm{k}_1}^{\ast})
(a_{\bm{k}_2} + a_{-\bm{k}_2}^{\ast})
(a_{\bm{k}_3} + a_{-\bm{k}_3}^{\ast})
}{\sqrt{\omega_{\bm{k}}\omega_{\bm{k}_1}\omega_{\bm{k}_2}\omega_{\bm{k}_3}}}
.
\end{align*}

It should be noted that
the companion mode for $a_{\bm{k}}$,
which has the wavenumber with the opposite sign $a_{-\bm{k}}^{\ast}$,
in the self-nonlinear interaction term of Eq.~(\ref{eq:FvKself})
appears in the first order
for consistency in the weakly nonlinear expansion.
In the weak turbulence systems that contain only the $2 \leftrightarrow 2$ interactions,
$\omega_{\bm{k}} + \omega_{\bm{k}}^{\mathrm{s}}$ is
the nonlinear frequency~\cite{newell01,Biven200128,Biven200398}.
However,
the companion mode must be taken into account
owing to the $1 \leftrightarrow 3$ and $3 \leftrightarrow 1$ interactions
in the nonlinear term in Eq.~(\ref{eq:fvka}).

Equation~(\ref{eq:FvKself}) can be rewritten in the following simultaneous equations:
\begin{align}
\frac{d}{dt}
\begin{pmatrix}
a_{\bm{k}} \\
a_{-\bm{k}}^{\ast}
\end{pmatrix}
 = 
&
\begin{pmatrix}
- i ( \omega_{\bm{k}} + \omega_{\bm{k}}^{\mathrm{s}} )
& -i \omega_{\bm{k}}^{\mathrm{s}}
\\
i \omega_{-\bm{k}}^{\mathrm{s}}
& i ( \omega_{-\bm{k}} + \omega_{-\bm{k}}^{\mathrm{s}} )
\end{pmatrix}
\begin{pmatrix}
a_{\bm{k}} \\
a_{-\bm{k}}^{\ast}
\end{pmatrix}
\nonumber\\
&
+
\begin{pmatrix}
\mathcal{N}_{\bm{k}}^{\mathrm{nn}} \\
{\mathcal{N}_{-\bm{k}}^{\mathrm{nn}}}^{\ast} \\
\end{pmatrix}
.
\label{eq:simultaneousa}
\end{align}
Because the extended self-nonlinear interactions preserve the inversion symmetry of the system,
the frequency increment satisfies the relation $\omega_{\bm{k}}^{\mathrm{s}} = \omega_{-\bm{k}}^{\mathrm{s}}$.
The eigenvalues of the matrix in the right-hand side of Eq.~(\ref{eq:simultaneousa})
are 
\begin{align}
\pm \omega_{\bm{k}}^{\mathrm{NL}}
 = \pm \omega_{\bm{k}} \sqrt{1 + \frac{2 \omega_{\bm{k}}^{\mathrm{s}}}{\omega_{\bm{k}}}}
.
\label{eq:defoNL}
\end{align}
If the remnant interaction term
$(\mathcal{N}_{\bm{k}}^{\mathrm{nn}}\ \ {\mathcal{N}_{-\bm{k}}^{\mathrm{nn}}}^{\ast})$
in Eq.~(\ref{eq:simultaneousa})
can be neglected, 
the solution is
\begin{align}
a_{\bm{k}}(t) =
&
 a_{\bm{k}}(0) 
\left( \cos \omega_{\bm{k}}^{\mathrm{NL}} t
 - i \frac{\omega_{\bm{k}} + \omega_{\bm{k}}^{\mathrm{s}}}{\omega_{\bm{k}}^{\mathrm{NL}}} \sin \omega_{\bm{k}}^{\mathrm{NL}} t
\right)
\nonumber\\
&
 - i 
a_{-\bm{k}}^{\ast}(0)
\frac{\omega_{\bm{k}}^{\mathrm{s}}}{\omega_{\bm{k}}^{\mathrm{NL}}} \sin \omega_{\bm{k}}^{\mathrm{NL}} t
,
\end{align}
which results in
\begin{align}
\zeta_{\bm{k}}(t) = \zeta_{\bm{k}}(0) \cos \omega_{\bm{k}}^{\mathrm{NL}} t .
\end{align}
In other words,
although $\omega_{\bm{k}}^{\mathrm{NL}} \approx \omega_{\bm{k}} + \omega_{\bm{k}}^{\mathrm{s}}$
for $\omega_{\bm{k}}^{\mathrm{s}} \ll \omega_{\bm{k}}$,
$\omega_{\bm{k}}^{\mathrm{NL}}$ is preferable 
to $\omega_{\bm{k}} + \omega_{\bm{k}}^{\mathrm{s}}$
which is considered as the nonlinear frequency 
in Refs.~\cite{newell01,Biven200128,Biven200398}.
The difference in the derivation of the nonlinear frequencies
in these references and in this paper
is whether the companion mode is taken into account.

It should be noted that
the evaluation of $\omega_{\bm{k}}^{\mathrm{s}}$ using Eq.~(\ref{eq:freqshift})
requires each Fourier component of the displacement $|\zeta_{\bm{k}}|^2$,
which is time-dependent in general.
However, because the frequencies must be constant in time to be meaningful,
the spectrum of the displacement should be replaced
by its ensemble- or time-average $\langle |\zeta_{\bm{k}}|^2 \rangle$.
In the statistically steady state,
the averaged $\omega_{\bm{k}}^{\mathrm{NL}} (\geq \omega_{\bm{k}})$
is expected to be a good approximation of the nonlinear frequency.

\subsection{Comparison with numerically obtained representative frequency}

\begin{figure*}
 \includegraphics[scale=1]{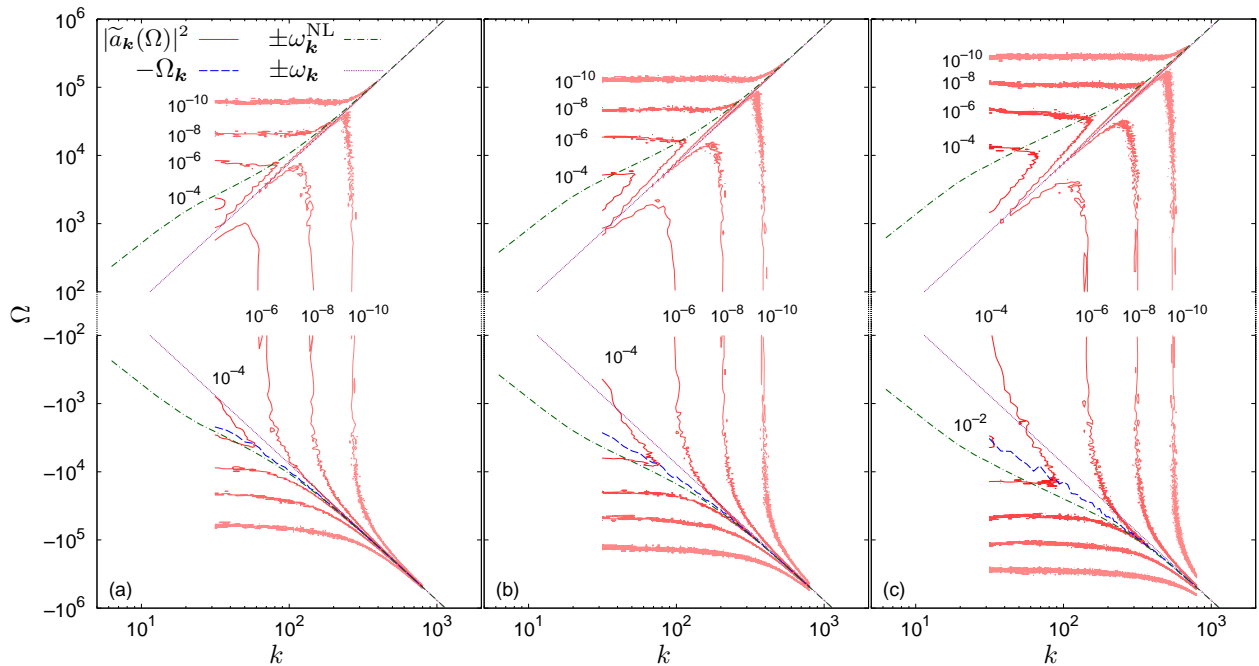}
\caption{(Color online)
 Contours of frequency spectra $|\widetilde{a}_{\bm{k}}(\Omega)|^2$
 for three energy levels. (a) EL1, (b) EL2, and (c) EL3.
 The contours are drawn with the (red) solid curves.
 The representative frequencies $\Omega_{\bm{k}}$,
 nonlinear frequencies $\omega_{\bm{k}}^{\mathrm{NL}}$,
 and  linear frequencies $\omega_{\bm{k}}$
 are, respectively, indicated by the (blue) broken,
 (green) dashed-dotted, and
 (purple) dotted curves.
\label{fig:freqsp}
}
\end{figure*}
The frequency spectrum $|\widetilde{a}_{\bm{k}}(\Omega)|^2$
is obtained from the Fourier transforms of the time series of $a_{\bm{k}}(t)$.
Note that $|\widetilde{a}_{\bm{k}}(\Omega)|^2$ lacks the symmetry
with respect to the sign of $\Omega$,
since $a_{\bm{k}}(t)$ is a complex number.
Figure~\ref{fig:freqsp} shows the contours
of the frequency spectra $|\widetilde{a}_{\bm{k}}(\Omega)|^2$.
The horizontal and vertical axes, which are scaled logarithmically, show the wavenumber and frequency, respectively.
Note that
the negative frequency is also scaled logarithmically,
and the figures for the positive and negative frequencies are arranged vertically
with gaps.
The contours (red curves) are drawn only in the inertial subrange $8\pi < k < 256\pi$. 
The cross-sections of these contours will exhibit a similar structure
to those in Fig.~3 of Ref.~\cite{PhysRevLett.110.105501}.
(See Appendix~\ref{sec:AppC} for the structure of the frequency spectrum.)

The {\em representative\/} nonlinear frequency $\Omega_{\bm{k}}$ for each $\bm{k}$ is defined
such that $|\widetilde{a}_{\bm{k}}(-\Omega_{\bm{k}})|^2$ is maximal.
The negative sign is introduced in the argument
because the maxima of $|\widetilde{a}_{\bm{k}}(\Omega)|^2$ 
always appear in the negative region of $\Omega$.
Therefore,
$\Omega=-\Omega_{\bm{k}}$
corresponds to the ridge curve of the contours of $|\widetilde{a}_{\bm{k}}(\Omega)|^2$.
Note that
$|\widetilde{a}_{\bm{k}}(\Omega)|^2 \propto \delta(\Omega + \omega_{\bm{k}})$
and $\Omega_{\bm{k}} = \omega_{\bm{k}}^{\mathrm{NL}} = \omega_{\bm{k}}$
if the nonlinear interactions were zero.
In Fig.~\ref{fig:freqsp},
the ridges of the frequency spectra are drawn only in the inertial subrange
with the (blue) broken curves.
The nonlinear frequencies $\omega_{\bm{k}}^{\mathrm{NL}}$,
which are obtained from Eqs.~(\ref{eq:freqshift}) and (\ref{eq:defoNL}) by averaging ten sets of data of $\zeta_{\bm{k}}$, 
as well as the linear frequencies~(\ref{eq:lineardispersion})
are drawn over the entire range of wavenumbers.

The structure of the contour curves drastically changes near
the linear dispersion line $\Omega=\pm \omega_{\bm{k}}$.
The asymptotic behavior of the contour curves in $|\Omega|<\omega_{\bm{k}}$ 
($|\Omega|>\omega_{\bm{k}}$) is vertical (horizontal),
which means that the spectral amplitude $|\widetilde{a}_{\bm{k}}(\Omega)|^2$
is nearly constant along the constant wavenumber (frequency) lines.
The hornlike extensions of the contour curves
to the upper right along $\omega_{\bm{k}}^{\mathrm{NL}}$
and to the lower right along $-\omega_{\bm{k}}^{\mathrm{NL}}$
correspond to the humps of $|\widetilde{a}_{\bm{k}}(\Omega)|^2$,
whereas the extension to the lower left along $\omega_{\bm{k}}$ corresponds to the depressions.
From the angles of the curves in Fig.~\ref{fig:freqsp}(a)--(c),
one can see that the humps and depressions broaden for higher energy levels.
The widths between the contour curves in Figs.~\ref{fig:freqsp}(a)--(c)
are also larger as the energy levels increase,
and it corresponds to a broadening of the frequency spectrum.

The negative frequency region is a primary interest to evaluate the nonlinear time scales.
For the large wavenumbers in the negative frequency region, where the nonlinearity is weak~\cite{PhysRevLett.110.105501},
the representative nonlinear frequencies $\Omega_{\bm{k}}$
are close to the linear frequencies $\omega_{\bm{k}}$
and the nonlinear frequencies $\omega_{\bm{k}}^{\mathrm{NL}}$.
Owing to the weak nonlinearity,
the frequency spectra are sharp around the maximum frequency,
and $\Omega_{\bm{k}}$ and $\omega_{\bm{k}}^{\mathrm{NL}}$ are barely distinguishable from $\omega_{\bm{k}}$.

For the small wavenumbers in the negative frequency region,
where the nonlinearity is relatively strong,
the frequency spectra are broad,
and the nonlinear frequencies $\omega_{\bm{k}}^{\mathrm{NL}}$
deviate from the linear frequencies $\omega_{\bm{k}}$ to the higher frequencies.
The representative nonlinear frequencies $\Omega_{\bm{k}}$
lie between $\omega_{\bm{k}}^{\mathrm{NL}}$ and $\omega_{\bm{k}}$.
Comparing the three figures, Figs.~\ref{fig:freqsp}(a)--(c),
we observe that
the difference among these three frequencies,
$\Omega_{\bm{k}}$, $\omega_{\bm{k}}$, and $\omega_{\bm{k}}^{\mathrm{NL}}$,
is larger for a wavenumber $k$ as the energy levels increase.

In the positive frequency regions shown in Fig.~\ref{fig:freqsp},
we found the secondary maxima of the frequency spectra
near the nonlinear frequencies $\omega_{\bm{k}}^{\mathrm{NL}}$
(see Appendix~\ref{sec:AppC}).
Our results show
the two eigenvalues~(\ref{eq:defoNL})
agree quite well with the primary maximum frequencies and the secondary maximum frequencies
in the large wavenumbers.
The secondary maxima of the frequency spectra
can be observed even in the weakly nonlinear regime,
which can be accounted for
by the coupling with the companion mode
via the $1 \leftrightarrow 3$ and $3 \leftrightarrow 1$ interactions
It is a different mechanism from the strong nonlinearity~\cite{zakharov_on_MMT}
to make the secondary maxima.

\begin{figure}[]
  \includegraphics[scale=.95]{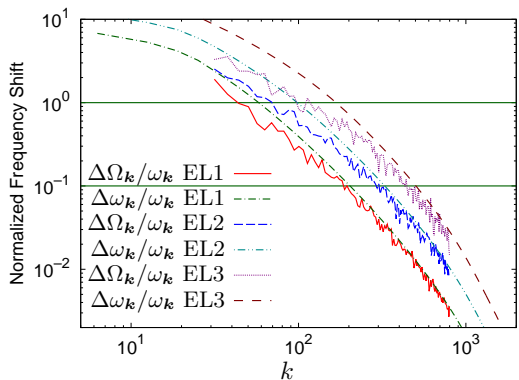}
 \caption{(Color online)
 Nonlinear frequency shifts normalized by the linear frequencies
 for three energy levels.
 The two horizontal lines represent the normalized frequency shifts equal to $10^{-1}$ and $1$.
 \label{fig:freqdeviation}
 }
\end{figure}
To observe the nonlinear frequency shifts more clearly,
the nonlinear frequency shifts normalized by the linear frequencies
are illustrated in Fig.~\ref{fig:freqdeviation}.
The frequency shifts due to the extended self-nonlinear interactions
and those of the representative frequency
are, respectively, obtained
as $\Delta \omega_{\bm{k}} = \omega_{\bm{k}}^{\mathrm{NL}} - \omega_{\bm{k}}$
and $\Delta \Omega_{\bm{k}} = \Omega_{\bm{k}} - \omega_{\bm{k}}$.
The frequency shifts $\Delta \omega_{\bm{k}}$
agree quite well with $\Delta \Omega_{\bm{k}}$
at low energies and large wavenumbers,
where the nonlinearity is weak.
Conversely,
at high energies and small wavenumbers,
the frequency shifts $\Delta \omega_{\bm{k}}$
overestimate the actual frequency shifts
since the remnant interactions
absorb the frequency shift due to the extended self-nonlinear interactions.
The nonlinear frequency 
$\omega_{\bm{k}}^{\mathrm{NL}} = \omega_{\bm{k}} \sqrt{1 + 2\omega_{\bm{k}}^{\mathrm{s}} / \omega_{\bm{k}}}$
agrees with $\Omega_{\bm{k}}$
better than $\omega_{\bm{k}} + \omega_{\bm{k}}^{\mathrm{s}}$
at small wavenumbers, at which $\omega_{\bm{k}}^{\mathrm{s}}$ is large.

The deviation of $\Delta \Omega_{\bm{k}}/\omega_{\bm{k}}$
from $\Delta \omega_{\bm{k}}/\omega_{\bm{k}}$
appears in the wavenumbers where $\Delta \Omega_{\bm{k}}/\omega_{\bm{k}}$
and $\Delta \omega_{\bm{k}}/\omega_{\bm{k}}$ are between $10^{-1}$ and $1$.
Since $\Delta \omega_{\bm{k}}$ is derived from the extended self-nonlinear interactions,
the deviation of $\Delta \Omega_{\bm{k}}/\omega_{\bm{k}}$
from $\Delta \omega_{\bm{k}}/\omega_{\bm{k}}$
is caused by the broad-mode interactions, 
including strongly nonlinear interactions.
Namely,
the deviation between $10^{-1}$ and $1$ indicates that
the strongly nonlinear effects are no longer negligible in this region.
Therefore,
the ratio $\Delta \omega_{\bm{k}}/\omega_{\bm{k}}$
can be understood as the measure of the nonlinearity for each wavenumber.
In other words,
the separation wavenumber between the weak and strong turbulence,
which is expected to be the application limit of the weak turbulence theory,
is determined by
the critical wavenumber where the ratio of the frequencies is between $10^{-1}$ and $1$.

\subsection{Identification of boundary wavenumber}
\label{ssec:cw}

The weak turbulence theory is not applicable
when the nonlinear time scale is comparable with the linear time scale.
Therefore,
the critical wavenumber $k_{\mathrm{c}}=|\bm{k}_{\mathrm{c}}|$ can be defined as 
\begin{align}
\Delta \omega_{\bm{k}_{\mathrm{c}}} =
\epsilon \omega_{\bm{k}_{\mathrm{c}}}
,
\label{eq:defepsilon}
\end{align}
where $\epsilon$ is a constant between $10^{-1}$ and $1$
as shown in Fig.~\ref{fig:freqdeviation}.

To find the analytical representation of the critical wavenumber $k_{\mathrm{c}}$ 
for arbitrary energy levels,
it is convenient to use the linear potential energy $\mathcal{V}(k)$.
As shown in the previous two subsections,
a good estimation of $\omega_{\bm{k}}^{\mathrm{NL}}$ 
as well as $\omega_{\bm{k}}^{\mathrm{s}}$ is obtained
by its ensemble- or time-average $\langle |\zeta_{\bm{k}}|^2 \rangle$.
Furthermore,
as derived in Appendix~\ref{sec:AppA},
$\langle |\zeta_{\bm{k}}|^2 \rangle \approx 4\pi \mathcal{V}(k) / (\rho k \omega_{\bm{k}}^2)$
in isotropic fields.
Therefore,
the averaged $\omega_{\bm{k}}^{\mathrm{NL}}$ is obtained from $\mathcal{V}(k)$.

\begin{figure}
  \includegraphics[scale=.9]{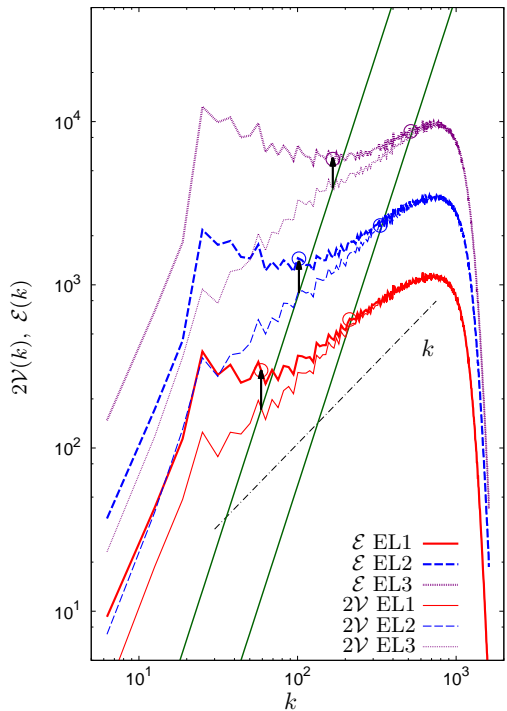}
 \caption{(Color online)
 Energy spectra (thick curves) and doubled linear potential energy spectra (thin curves)
 for three energy levels.
 The left and right inclined parallel lines represent $2\Check{\mathcal{V}}(k_{\mathrm{c}}; 1)$ and $2\Check{\mathcal{V}}(k_{\mathrm{c}}; 10^{-1})$, respectively.
The circles indicate the position of the energy spectra at the wavenumbers
at which the linear potential energy spectra intersect with $2\Check{\mathcal{V}}(k_{\mathrm{c}}; 1)$ and with $2\Check{\mathcal{V}}(k_{\mathrm{c}}; 10^{-1})$.
 \label{fig:energyspectra}
 }
\end{figure}

It is assumed that the linear potential energy has the self-similar form $\mathcal{V}(k)=C k$.
The self-similarity of $\mathcal{V}(k)$ over the entire inertial subrange for all the energy levels
will be seen in Fig.~\ref{fig:energyspectra},
which is a distinctive feature from the energy spectra $\mathcal{E}(k)$.
Equations~(\ref{eq:freqshift}), (\ref{eq:defoNL}), and (\ref{eq:defepsilon}) 
give the analytical representation of the critical wavenumber
with the parameter $C$ corresponding to the energy level
as
\begin{align}
k_{\mathrm{c}} = \frac{6(1-\sigma^2)}{h^2} \sqrt{\frac{3C}{\epsilon(\epsilon+2)E}}
\label{eq:kc-estmt}
\end{align}
(see Appendix~\ref{sec:AppB} for the derivation).

There exists a more convenient representation
to graphically find the value of $k_{\mathrm{c}}$.
Equation~(\ref{eq:kc-estmt}) can be rewritten as
\begin{align}
\mathcal{V}(k_{\mathrm{c}})
 = \frac{\epsilon(\epsilon+2)Eh^4}{108(1-\sigma^2)^2}k_{\mathrm{c}}^3
\equiv \Check{\mathcal{V}}(k_{\mathrm{c}}; \epsilon)
.
\label{eq:Vkc}
\end{align}
To illustrate the validity of Eq.~(\ref{eq:Vkc}),
the energy spectra $\mathcal{E}(k)$ and twice the linear potential energy spectra $2\mathcal{V}(k)$
are drawn in Fig.~\ref{fig:energyspectra}.
The two inclined parallel lines in Fig.~\ref{fig:energyspectra}
represent $2\Check{\mathcal{V}}(k_{\mathrm{c}}; 10^{-1})$ and $2\Check{\mathcal{V}}(k_{\mathrm{c}}; 1)$.
The critical wavenumber $k_{\mathrm{c}}$ for $\epsilon$ can be found
at the intersection of $2\mathcal{V}(k)$ and $2\Check{\mathcal{V}}(k_{\mathrm{c}}; \epsilon)$.
Then,
our estimation predicts
the separation wavenumber is located
between the critical wavenumber for $\epsilon=10^{-1}$ and that for $\epsilon=1$.
In other words, the wavenumbers at which $\mathcal{E}(k)$'s bend
are located between $k_{\mathrm{c}}$ for $\epsilon=10^{-1}$ and that for $\epsilon=1$
for all the energy levels.
At the larger wavenumber side
of the line $2\Check{\mathcal{V}}(k_{\mathrm{c}}; 10^{-1})$,
we find that $\mathcal{E}(k) \approx 2\mathcal{V}(k)$
and that the system is weakly nonlinear.
Conversely,
at the smaller wavenumber side
of the line $2\Check{\mathcal{V}}(k_{\mathrm{c}}; 1)$,
we found that $\mathcal{E}(k)$ is far from $2\mathcal{V}(k)$
and that $a_{\bm{k}}$ cannot be approximated by the harmonic oscillation.

\begin{figure}
 \includegraphics[scale=.9]{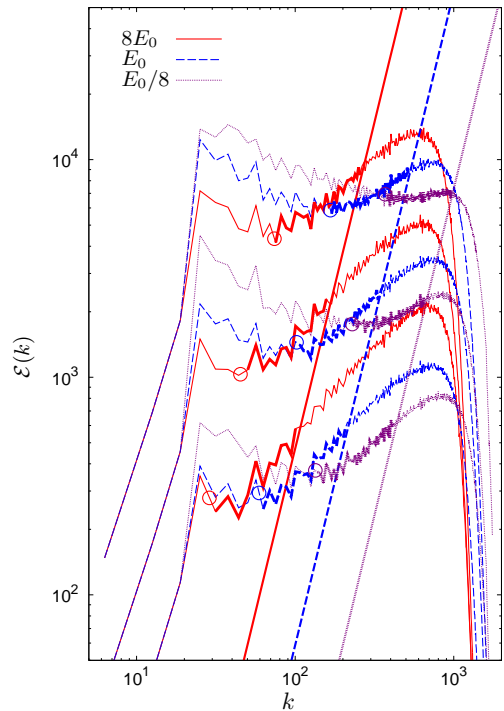}
 \caption{(Color online)
Energy spectra for different values of the Young's modulus.
The solid lines show $2\Check{\mathcal{V}}(k_{\mathrm{c}}; 10^{-1})$ for each Young's modulus.
The energy spectra between $k_{\mathrm{c}}$ for $\epsilon=10^{-1}$ and that for $\epsilon=1$
are indicated by the thick curves,
and the circles indicate $\mathcal{E}(k_{\mathrm{c}})$
where $k_{\mathrm{c}}$ is determined by $\mathcal{V}(k_{\mathrm{c}})=\Check{\mathcal{V}}(k_{\mathrm{c}}; 1)$.
\label{fig:spectra4E}}
\end{figure}

To verify Eq.~(\ref{eq:Vkc}),
other series of DNS, in which the material parameters
such as the Young's modulus $E$ and thickness $h$ 
of the elastic plates are artificially changed,
were performed.
The energy spectra for different values of the Young's modulus are shown
in Fig.~\ref{fig:spectra4E}.
Because we used $8E_0$ and $E_0/8$ for the Young's modulus,
the critical wavenumbers should be $2k_{\mathrm{c}0}$ and $k_{\mathrm{c}0}/2$
according to Eq.~(\ref{eq:Vkc}),
where $E_0$ is the Young's modulus of the originally used steel plate
and $k_{\mathrm{c}0}$ is the critical wavenumber for $E=E_0$.
Clearly,
Fig.~\ref{fig:spectra4E} supports that
the right-hand side in Eq.~(\ref{eq:Vkc}) is proportional to $E k_{\mathrm{c}}^3$.

\begin{figure}
 \includegraphics[scale=.9]{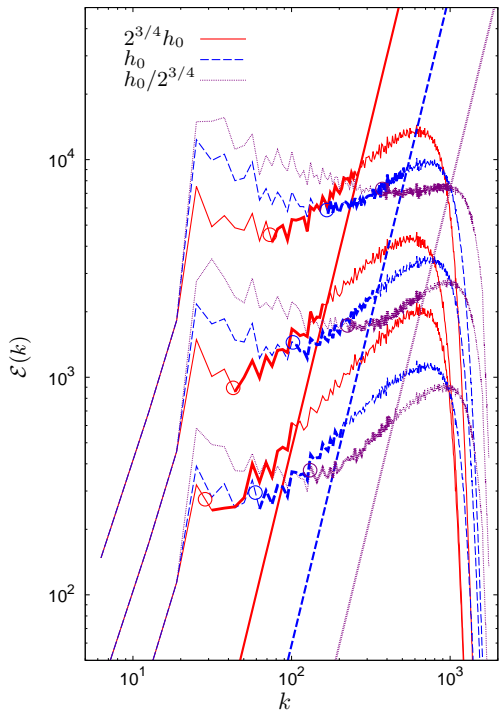}
 \caption{(Color online)
Energy spectra for different values of the thickness of the plate.
The solid lines show $2\Check{\mathcal{V}}(k_{\mathrm{c}}; 10^{-1})$ for each thickness.
The energy spectra between $k_{\mathrm{c}}$ for $\epsilon=10^{-1}$ and that for $\epsilon=1$
are indicated by the thick curves,
and the circles indicate $\mathcal{E}(k_{\mathrm{c}})$
where $k_{\mathrm{c}}$ is determined by $\mathcal{V}(k_{\mathrm{c}})=\Check{\mathcal{V}}(k_{\mathrm{c}}; 1)$.
\label{fig:spectra4h}}
\end{figure}

Similarly,
the energy spectra for different values of the thickness $h$
are shown in Fig.~\ref{fig:spectra4h}.
Because we used $2^{3/4} h_0$ and $h_0/2^{3/4}$ for the thickness,
the critical wavenumbers should be $2k_{\mathrm{c}0}$ and $k_{\mathrm{c}0}/2$
again,
where $h_0$ is the originally used thickness.
This also supports that
the coefficient in Eq.~(\ref{eq:Vkc}) is proportional to $h^4 k_{\mathrm{c}}^3$.
We also confirmed that
the critical wavenumber does not depend on the density $\rho$,
although the figures are omitted here.
Note that we do not examine the dependence of the critical wavenumber on the Poisson ratio $\sigma$
since the variation of the Poisson ratio $\sigma$ among elastic media is small.
In Figs.~\ref{fig:energyspectra}, \ref{fig:spectra4E}, and \ref{fig:spectra4h},
we have demonstrated the validity of the estimation of the separation wavenumber
according to Eq.~(\ref{eq:Vkc}) including its coefficients.

\section{Conclusion}

In this paper,
we have obtained the analytical representation of the critical wavenumbers
at which the nonlinear frequency shifts are comparable with the linear frequencies.
In the derivation of the nonlinear frequency,
the extended self-nonlinear interactions including the coupling with the companion elementary wave
are explicitly evaluated.
The coupling with the companion elementary wave must be taken into account
generally in the weak turbulence systems
containing the $1 \leftrightarrow 3$ and $3 \leftrightarrow 1$ interactions as well as the $2 \leftrightarrow 2$ interactions.
The nonlinear frequencies evaluated 
by the extended self-nonlinear interactions
agree well with the representative nonlinear frequencies
determined by the maxima of the frequency spectra obtained 
in the direct numerical simulations. 
The agreement is remarkable at large wavenumbers,
at which the weak turbulence spectrum is observed.

The analytically obtained critical wavenumber successfully reproduced 
the separation wavenumber between the weak and strong turbulence spectra.
It is consistent with the view that
the weak turbulence theory is not applicable for wavenumbers
at which the nonlinear time scales are comparable with the linear time scales.
The agreement between the critical wavenumbers and the separation wavenumbers
as well as the dependence of the wavenumbers on the material parameters
were also confirmed.

The coexistence of the weak and strong turbulence spectra
implies that the large-scale strongly nonlinear coherent structures
coexist with a large number of small-scale weakly nonlinear random waves
in the real space.
The characterization of each structure is of great importance.
The relations of this real-space coexistence
to the different properties between $\mathcal{E}(k)$ and $\mathcal{V}(k)$
appearing in Fig.~\ref{fig:energyspectra}
are now under consideration.
We will also identify the energy transfer among waves in the coexistence.

\appendix
\section{Structures of frequency spectra}
\label{sec:AppC}

\begin{figure}[]
 \includegraphics[scale=.95]{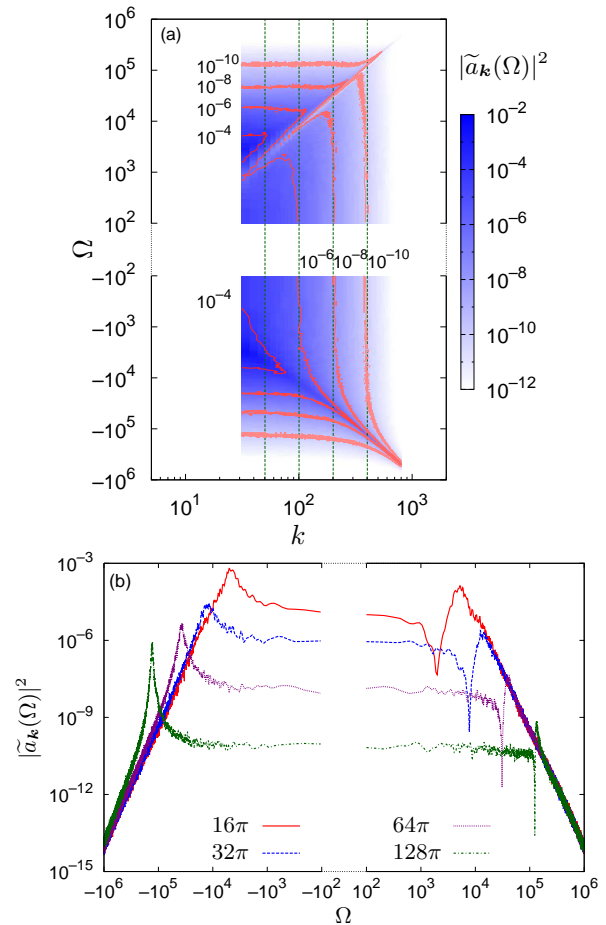}
\caption{(Color online)
(a): Density plot and contours of the frequency spectra $|\widetilde{a}_{\bm{k}}(\Omega)|^2$ for EL2.
 The vertical solid lines show $k=16\pi$, $32\pi$, $64\pi$, and $128\pi$.
(b): Frequency spectra for $k=16\pi$, $32\pi$, $64\pi$, and $128\pi$ for EL2.
The positive and negative frequencies are arranged horizontally with the gap.
\label{fig:freq2}
}
\end{figure}

The density plot and contours of the frequency spectra for EL2 are shown
in Fig.~\ref{fig:freq2}(a).
It corresponds to Fig.~\ref{fig:freqsp}(b),
where only the contours are drawn for the spectra.
The ridge of the frequency spectra in the negative frequencies
corresponds to the cusps of the contours which extend to the lower right.
Similarly,
the ridge and trough in the positive frequencies
correspond to the cusps which extend to the upper right and lower left, respectively.
To see the ridges and trough more clearly,
the frequency spectra for $k=16\pi$, $32\pi$, $64\pi$, and $128\pi$
on the line parallel to the vertical axis at each $k$
are shown in Fig.~\ref{fig:freq2}(b).
Humps can be observed in the positive and negative frequency regions of $\Omega$;
the hump in the negative frequency region is larger than that in the positive frequency region.
The null amplitude $|\widetilde{a}_{\bm{k}}(\omega_{\bm{k}})|^2 = 0$
produces a slit in the right humps.
This structure corresponds to the hornlike extension of the contour curves to the lower right
along $\omega_{\bm{k}}$ in Fig.~\ref{fig:freqsp}.
The secondary maximum frequencies are located near $\omega_{\bm{k}}^{\mathrm{NL}}$.

The $|\widetilde{a}_{\bm{k}}(\omega_{\bm{k}})|^2 = 0$ property
is a direct consequence of the definition of $a_{\bm{k}}$.
The frequency-space representation of Eq.~(\ref{eq:fvkzetapb}),
$\widetilde{p}_{\bm{k}}(\Omega) = i \rho \Omega \widetilde{\zeta}_{\bm{k}}(\Omega)$,
and the definition of the complex amplitude, Eq.~(\ref{eq:defa}),
lead to the relation $|\widetilde{a}_{\bm{k}}(\Omega)|^2 = \rho (\omega_{\bm{k}}-\Omega)^2|\widetilde{\zeta}_{\bm{k}}(\Omega)|^2 /(2\omega_{\bm{k}})$.
We have checked these relations in our simulations,
though the figures are omitted here.

\section{Spectra of linear potential energy and of displacement}
\label{sec:AppA}

In the present study,
the periodic boundary condition with a period of $L=1$[m] is used.
The Fourier series is defined as
\begin{align*}
 F_{\bm{k}} &= \frac{1}{L^2} \int_0^L\int_0^L d\bm{x} f(\bm{x}) \exp(-i \bm{k} \cdot \bm{x})
,
 \nonumber\\
 f(\bm{x})  &= \sum_{\bm{k}} F_{\bm{k}} \exp(i \bm{k} \cdot \bm{x})
,
\end{align*}
where $f(\bm{x})$ is a physical quantity such as the displacement $\zeta(\bm{x})$ and momentum $p(\bm{x})$ in the real space,
and $F_{\bm{k}}$ is its Fourier coefficient.
The azimuthally integrated linear potential energy spectrum is given as
\begin{align*}
 \mathcal{V}(k) \Delta k &= \sum_{k-\Delta k/2 \leq |\bm{k}| < k+\Delta k/2}
 \frac{\rho\omega_{\bm{k}}^2}{2} |\zeta_{\bm{k}}|^2
,
\end{align*}
where $\Delta k$ is the bin width used to obtain the spectrum;
$\Delta k$ is set to $2\pi/L$
in the present study.
The summation
in the two-dimensional wavenumber space
can be replaced by the integral as
\begin{align*}
 \sum_{\bm{k}} \approx \int_{0}^{\infty} dk \int_{0}^{2\pi} d\theta  \frac{k}{\Delta k_x \Delta k_y}
.
\end{align*}

In a statistically isotropic system,
the ensemble- and/or time-averaging $\langle \cdots \rangle$ is performed
to evaluate the spectra:
\begin{align*}
\mathcal{V}(k) \Delta k
 & \approx \frac{2\pi k \Delta k}{\Delta k_x \Delta k_y}
 \frac{\rho\omega_{\bm{k}}^2}{2} \langle |\zeta_{\bm{k}}|^2 \rangle
,
\end{align*}
where $\Delta k_x = \Delta k_y = 2\pi/L$ is the grid spacing.

\section{Frequency due to self-interactions and linear potential energy spectra}
\label{sec:AppB}

In an isotropic system,
the frequency due to the self-interactions
can be rewritten as
\begin{align}
\omega_{\bm{k}}^{\mathrm{s}} 
 &= \frac{E}{2\pi \rho^2 \omega_{\bm{k}}}
 \int_{0}^{\infty} dk^{\prime} \int_{0}^{2\pi} d\theta^{\prime}
\frac{|\bm{k} \times \bm{k}^{\prime}|^4}{|\bm{k} - \bm{k}^{\prime}|^4}
\frac{\mathcal{V}(k^{\prime})}{\omega_{\bm{k}^{\prime}}^2}
\nonumber
\\
 &= \frac{3Ek^4}{8 \rho^2 \omega_{\bm{k}}^3}
\left(
 \int_0^k dk^{\prime} \mathcal{V}(k^{\prime})
+ k^4 \int_k^{\infty} dk^{\prime} k^{\prime -4} \mathcal{V}(k^{\prime})
\right)
,
\label{eq:doE}
\end{align}
where the integral over $\theta^{\prime}$ is performed
using the following equation for a constant $c$:
\begin{align*}
 \int_{0}^{2\pi} d\theta \frac{\sin^4 \theta}{(1 + c^2 + 2c\cos\theta)^2}
 = 
\begin{cases}
\displaystyle\frac{3\pi}{4} & \text{for\ } 0 \leq c \leq 1
\\
\displaystyle\frac{3\pi}{4c^4} & \text{for\ } c>1 
\end{cases}
.
\end{align*}

The self-similarity of the linear potential energy is assumed
as $\mathcal{V}(k) = C k^{\alpha}$.
Then,
the two integrals in Eq.~(\ref{eq:doE})
converge when $-1 < \alpha<3$,
and Eq.~(\ref{eq:doE}) is reduced to
\begin{align}
\omega_{\bm{k}}^{\mathrm{s}}
 &= 
\frac{3}{2(\alpha+1)(3-\alpha)}
\sqrt{\frac{12^3 (1-\sigma^2)^3}{\rho Eh^6}}
\frac{\mathcal{V}(k)}{k}
.\label{eq:oksV}
\end{align}
Finally,
Eqs.~(\ref{eq:defoNL}), (\ref{eq:defepsilon}), and (\ref{eq:oksV}) yield Eqs.~(\ref{eq:kc-estmt}) and (\ref{eq:Vkc})
for $\alpha=1$.

\begin{acknowledgments}
The authors thank an anonymous referee for suggesting recently published papers.
This work was partially supported by KAKENHI Grant No.~25400412.
\end{acknowledgments}

\end{document}